\documentclass[12pt,epsf]{article}
\usepackage{graphicx}
\usepackage{amssymb}
\usepackage{color}

\begin{document}

\def\a{\alpha}
\def\b{\beta}
\def\c{\varepsilon}
\def\d{\delta}
\def\e{\epsilon}
\def\f{\phi}
\def\g{\gamma}
\def\h{\theta}
\def\k{\kappa}
\def\l{\lambda}
\def\m{\mu}
\def\n{\nu}
\def\p{\psi}
\def\q{\partial}
\def\r{\rho}
\def\s{\sigma}
\def\t{\tau}
\def\u{\upsilon}
\def\v{\varphi}
\def\w{\omega}
\def\x{\xi}
\def\y{\eta}
\def\z{\zeta}
\def\D{\Delta}
\def\G{\Gamma}
\def\H{\Theta}
\def\L{\Lambda}
\def\F{\Phi}
\def\P{\Psi}
\def\S{\Sigma}
\def\BR{{\rm Br}}
\def\o{\over}
\def\beq{\begin{eqnarray}}
\def\eeq{\end{eqnarray}}
\newcommand{\nn}{\nonumber \\}
\newcommand{\gsim}{ \mathop{}_{\textstyle \sim}^{\textstyle >} }
\newcommand{\lsim}{ \mathop{}_{\textstyle \sim}^{\textstyle <} }
\newcommand{\vev}[1]{ \left\langle {#1} \right\rangle }
\newcommand{\bra}[1]{ \langle {#1} | }
\newcommand{\ket}[1]{ | {#1} \rangle }
\newcommand{\EV}{ {\rm eV} }
\newcommand{\KEV}{ {\rm keV} }
\newcommand{\MEV}{ {\rm MeV} }
\newcommand{\GEV}{ {\rm GeV} }
\newcommand{\TEV}{ {\rm TeV} }
\def\diag{\mathop{\rm diag}\nolimits}
\def\Spin{\mathop{\rm Spin}}
\def\SO{\mathop{\rm SO}}
\def\O{\mathop{\rm O}}
\def\SU{\mathop{\rm SU}}
\def\U{\mathop{\rm U}}
\def\Sp{\mathop{\rm Sp}}
\def\SL{\mathop{\rm SL}}
\def\tr{\mathop{\rm tr}}

\newcommand{\bear}{\begin{array}}  
\newcommand {\eear}{\end{array}}
\newcommand{\la}{\left\langle}  
\newcommand{\ra}{\right\rangle}
\newcommand{\non}{\nonumber}  
\newcommand{\ds}{\displaystyle}
\newcommand{\red}{\textcolor{red}}
\def\ubl{U(1)$_{\rm B-L}$}
\def\REF#1{(\ref{#1})}
\def\lrf#1#2{ \left(\frac{#1}{#2}\right)}
\def\lrfp#1#2#3{ \left(\frac{#1}{#2} \right)^{#3}}
\def\OG#1{ {\cal O}(#1){\rm\,GeV}}


\baselineskip 0.7cm

\begin{titlepage}

\vskip 1.35cm
\begin{center}
{\large \bf
Large CP Violation in $B_s$ Meson Mixing \\ 
with EDM constraint in Supersymmetry
}
\vskip 1.2cm
Motoi Endo$^{1,2}$ and Norimi Yokozaki$^1$
\vskip 0.4cm

{\it $^1$ Department of Physics, University of Tokyo,
   Tokyo 113-0033, Japan\\
$^2$ Institute for the Physics and Mathematics of the Universe, 
University of Tokyo,\\ Chiba 277-8568, Japan
}

\vskip 1.5cm

\abstract{
Motivated by the recent measurement of the like-sign dimuon charge asymmetry, 
we investigate anomalous CP violation in the $B_s-\bar{B}_s$ mixing within the supersymmetry. 
We show that when gluino diagrams dominate supersymmetry contributions, it is very difficult to 
realize a large $B_s-\bar{B}_s$ mixing phase under the constraint from electric dipole moments 
barring cancellations. This constraint can be ameliorated by supposing superparticles decoupled. 
In this limit, we find that it is possible to achieve the large CP asymmetry, and the branching 
ratio of $B_s \to \mu^+ \mu^-$ tends to become sizable.
}
\end{center}
\end{titlepage}

\setcounter{page}{2}

\section{Introduction}

A lot of experimental efforts have confirmed the violation of the CP symmetry in the quark sector 
such as in the $K$ and $B$ mesons. Although the CP violation is sensitive to physics beyond the 
standard model (SM), most of the CP-violating phenomena are well explained within the framework 
of the SM. In spite of the experimental constraints, a room is still left for the $b-s$ mixing sector~\cite{Gabbiani:1996hi}.

The Tevatron experiments study possible effects of the CP violation in the $B$ system.  
The D0 collaborations recently reported a measurement of the like-sign dimuon charge 
asymmetry. Interpreting the result by the mixing of the neutral $B$ mesons, we obtain 
$(A_{sl}^b)_{\rm exp} = -(9.57\pm 2.51 \pm 1.46)\times 10^{-3}$~\cite{Abazov:2010hv,Abazov:2010hj}. 
Based on the Tevatron data, the asymmetry is also expressed in terms of the semileptonic 
asymmetry as $A_{sl}^b \simeq 0.5 a_{sl}^d + 0.5 a_{sl}^s$~\cite{Abazov:2010hv,Abazov:2010hj}. 
Noting that the semileptonic asymmetry is related 
to the $B$ meson mixing, the dimuon charge asymmetry is predicted by the SM as 
$(A_{sl}^b)_{\rm SM} = (-2.3_{-0.6}^{+0.5}) \times 10^{-4}$~\cite{Lenz:2006hd}. 
Interestingly, the experimental result is larger than the SM prediction at the 3.2$\sigma$ level. 
Combined with the other experimental data which are sensitive to the $B-\bar B$ mixing, the 
dimuon charge asymmetry favors an anomolous CP phase arising in the $B_s$ meson 
mixing~\cite{Ligeti:2010ia}.  

As a new source of the CP violation, the supersymmetric (SUSY) extension of the SM involves 
soft  SUSY breaking parameters. In particular, CP violating phases generally give arise in 
flavor-changing components of the soft scalar masses. They contribute to flavor-changing 
neutral currents (FCNCs) with CP violations. Actually, it has been argued that CP-violating $b-s$ 
squark mixings are a suitable candidate for the anomalous CP violation which is favored by the 
like-sign dimuon charge asymmetry~\cite{Parry:2010ce,Ko:2010mn,King:2010np}.

However, flavor and CP violations have been restricted experimentally. Especially
the electric dipole moments (EDMs) are sensitive to anomalous CP violations. It was pointed 
out that CP-violating $b-s$ squark mixings induce the chromo electric dipole moment (CEDM) of 
the strange quark and are tightly constrained by the atomic and neutron EDMs~\cite{Hisano:2003iw}. 
Nevertheless, this constraint has been often discarded in previous works on CP-violating FCNCs, 
because estimation of EDMs includes potentially large hadronic uncertainties and may also be 
suppressed by (accidental) cancellations. In contrast, we investigate the $B_s-\bar B_s$ mixing 
phase with seriously taking account of the EDM constraint. We will show that it is very difficult 
to realize such a large CP asymmetry unless we assume cancellations among the SUSY contributions.

A large CP violation may be reconciled with the EDM bound by supposing cancellations among 
SUSY contributions to the EDMs. This may be indeed a possible solution, because the phase 
structure of the SUSY contributions to the $B_s - \bar B_s$ mixing differs from that of the 
EDMs~\cite{Endo:2010fk}. However, since the EDM bound is too tight, the cancellation is quite 
easily upset by radiative corrections unless they are protected by symmetries.

Apart from the cancellations, a simple solution to avoid the EDM bound is to decouple the (colored) 
superparticles. Heavy superparticles suppresses most of the SUSY contributions to FCNCs as well 
as those to the EDMs. In this limit, the FCNCs can be induced by mediating the heavy Higgs bosons 
at the BLO level~\cite{Isidori:2001fv}. Actually, non-holomorphic (flavor-changing) interactions can be 
left sizable in the decoupling limit~\cite{Hall:1993gn,Carena:1994bv}. We will show that there is a 
wide parameter region which realizes the large CP violation in the $B_s - \bar B_s$ mixing with 
alleviating the EDM constraints. In addition, we will predict a relatively large branching ratio of the 
$B_s \to \mu\mu$ decay, and it is expected to be measured in the LHCb experiment.

\section{CP Phase in $B_s$ Meson Mixing}

The like-sign dimuon charge asymmetry implies an anomalous CP-violating contribution to 
the $B$ meson mixing~\cite{Abazov:2010hv,Abazov:2010hj}. The asymmetry is expressed 
by the $B_q -\bar{B}_q$ mixing parameters through the semileptonic asymmetry, 
$a_{sl}^{q} = {\rm Im} (\Gamma_{12}^{q}/M_{12}^{q})$, where $M_{12}^{q}$ and 
$\Gamma_{12}^q$ are the dispersive and the absorptive part of the $B_q -\bar{B}_q$ mixing 
amplitudes, respectively. On the other hand, the parameters also contribute to the other 
processes, e.g.~$\Delta M_q$, $\Delta \Gamma_q$ and time-dependent CP asymmetries 
of the $B_q$ meson decays. According to the global fit analysis in Ref.~\cite{Lenz:2006hd}, 
the authors searched for the $\chi^2$ minimum for the processes which are sensitive to the 
$B_q$ meson mixings as well as the like-sign dimuon charge asymmetry. Despite that the 
SM predicts a tiny angle as $\arg(\Gamma_{12}^{s}/M_{12}^{s}) \sim 10^{-(2-3)}$, the result 
indicates an anomalously large CP phase of $O(0.1-1)$ in the $B_s-\bar B_s$ mixing~\cite{Ligeti:2010ia}. 
\footnote{ The measurements of the $B_s \to J/\psi \phi$ decay have been recently updated 
by the Tevatron experiments~\cite{CDFNOTE,D0NOTE}. Although the results are turned to be 
consistent with the SM prediction, the experimental uncertainty is still large. }
The global fit result also prefers a larger absorptive part of the $B_s -\bar{B}_s$ mixing than 
the SM prediction. However, both the experimental and SM values are largely 
uncertain~\cite{Ligeti:2010ia,Bauer:2010dga}. Thus, we focus on new physics (NP) contributions 
to the $B_s - \bar B_s$ mixing phase.

In order to investigate the NP contribution to the $B_s - \bar{B}_s$ mixing, we parametrize 
dispersive part of the $B_s -\bar{B}_s$ mixing amplitude as 
\begin{equation}
 M_{12}^s = (M_{12}^s)^{\rm SM} + (M_{12}^s)^{\rm NP} 
 = (M_{12}^s)^{\rm SM}(1+h_s e^{2i\sigma_s}),
\end{equation}
where $(M_{12}^s)^{\rm SM}$ and $(M_{12}^s)^{\rm NP}$ are the SM and NP contributions, 
respectively. This contributes to the mass difference of the neutral $B_s$ mesons, $\Delta M_s$, 
as well as the semileptonic asymmetry, $a_{sl}^{q}$. Since the measurements of $\Delta M_s$ 
at the Tevatron are consistent with the SM prediction~\cite{Abazov:2006dm,Abulencia:2006ze}, 
$(M_{12}^s)^{\rm NP}$ is restricted as $|1 + h_s e^{2i\sigma_s}| \simeq 1$. On the other hand, 
$\sigma_s$ is favored to be large by the measurement of the like-sign dimuon charge asymmetry. 
Thus, $h_s$ must be correlated with $\sigma_s$ for $\sigma_s \sim 1$. In the absence of the NP 
contributions to the $B_d-\bar B_d$ mixing, the $\chi^2$ minimum is found around~\cite{Ligeti:2010ia}
\begin{equation}
 (h_s, \sigma_s) \simeq (1.8, 100^{\circ}).
 \label{eq:dimuon}
\end{equation}
We will refer this value in the following discussion. Note that the following result does not change 
qualitatively as long as the experimental data favor a large mixing angle even if they shift in future.

\section{Gluino Dominant Case}

\begin{figure}[t!]
\begin{center}
\includegraphics[width=7.5cm]{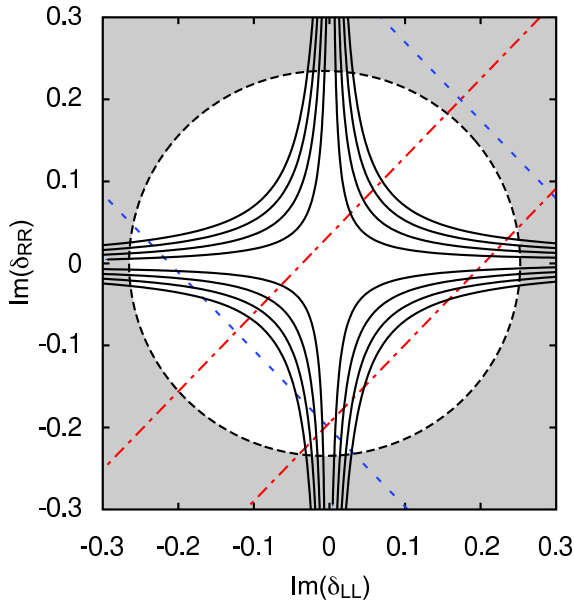}
\end{center}
\caption{
Contours of $h_s = 2.0, 1.5, 1.0$ and 0.5 from the outside to inside (solid line). The axes are the 
imaginary component of the $b-s$ squark mixings. 
The inside of the dashed line (circle) is allowed by ${\rm Br}(b\to s\gamma)$ at the 2$\sigma$ level. 
The region between the dotted-dashed (red) lines is the 2$\sigma$ band of $S_{\eta' K}$, and that 
inside the dotted (blue) lines is of $S_{\phi K}$. We set $m_{\rm SUSY}=\mu=500\GEV$, ${\rm Re}
(\delta^d_{LL})_{23}={\rm Re}(\delta^d_{RR})_{23}=0$ and $\tan\beta=10$.}
\label{fig:small_msoft}
\end{figure}

When superparticles have a mass of $O(100-1000)$GeV, gluino--squark diagrams usually
dominate SUSY contributions to the colored FCNC and CP violating processes. 
In the super-CKM basis~\cite{Gabbiani:1996hi}, the down- and up-type squark mass matrices 
are described as
\begin{equation}
 (M_{\tilde d}^2)_{ij} = {\rm diag}(m_{\tilde{d}}^2)_{ii} + m_{\tilde{d}}^2 (\delta^d)_{ij},~~~  
 (M_{\tilde u}^2)_{ij} = {\rm diag}(m_{\tilde{u}}^2)_{ii} + m_{\tilde{u}}^2 (\delta^u)_{ij},
\end{equation}
where the first term in the right-hand side denotes the squark mass, and the second one 
represents the flavor-changing components with an average squark mass $m_{\tilde{q}}^2$. 
Depending on the chirality structure of the squarks, the squark mixings are classified to the 
four types, $(\delta^q_{LL})_{ij}$, $(\delta^q_{LR})_{ij}$, $(\delta^q_{RL})_{ij}$ and 
$(\delta^q_{RR})_{ij}$, with $q = d, u$ and $i, j$ the generation index. Then, the SUSY 
contributions to the CP-violating FCNCs are represented by the squark mixings and the masses 
of the squarks and gluino. It is noticed that the complex phase of the SUSY contributions is not 
rephasing invariant. We choose the phase convention such that the SM contributions are real, 
if not otherwise mentioned. Namely, the phase of $(\delta^q)_{ij}$ represents the CP violation.

The SUSY contribution to $M_{12}^s$ is dominated by the gluino box diagrams. 
Normalizing it by the SM contribution, $h_s e^{2i\sigma_s}$ is evaluated as~\cite{Gabbiani:1996hi}
\begin{eqnarray}
  h_s e^{2i\sigma_s} &\simeq& 
  a_1 \left[ (\delta^d_{LL})_{23}^2 + (\delta^d_{RR})_{23}^2 \right]
  - a_2 \left[ (\delta^d_{LL})_{23} (\delta^d_{RR})_{23} \right]
  \nonumber \\
  && + a_3 \left[ (\delta^d_{LR})_{23}^2 + (\delta^d_{RL})_{23}^2 \right]
  - a_4 \left[ (\delta^d_{LR})_{23} (\delta^d_{RL})_{23} \right],
  \label{eq:b-bbar}
\end{eqnarray}
where the coefficients depend on the detail of the mass spectrum of the squarks and gluino, whose 
size is $a_1 = O(1), a_2 = O(100), a_3 = O(10)$ and $a_4 = O(10)$ for $m_{\tilde q} \sim m_{\tilde g} 
\sim 500$GeV (see e.g.~\cite{Gabbiani:1996hi} for details). Let us mention that they are scaled as 
$a_i \propto (\delta^d)_{23}^2 m_{\rm soft}^{-2}$, while it is insensitive to $\tan\beta$ when the gluino 
diagram is dominant. On the other hand, it is noted that the second line of (\ref{eq:b-bbar}) consists 
of the squark mixings with chirality flips. Since they are tightly constrained by ${\rm Br}(b \to s\gamma)$ 
as $(\delta^d_{LR,RL})_{23} < O(10^{-2})$ for $m_{\tilde q}, m_{\tilde g} \sim 500$GeV~\cite{Gabbiani:1996hi}, their contributions are negligible. Thus, we assume $(\delta^d_{LR,RL})_{23} = 0$ for simplicity. 

We show contours of $h_s$ in Fig.~\ref{fig:small_msoft}, where we set $m_{\tilde q} = m_{\tilde g} 
= 500$GeV and ${\rm Re}(\delta^d_{LL})_{23}={\rm Re}(\delta^d_{RR})_{23}=0$. It is stressed that 
$h_s$ is enhanced by a combination of $(\delta^d_{LL})_{23}$ and $(\delta^d_{RR})_{23}$ as is 
found in (\ref{eq:b-bbar}). Thus, it becomes comparable to the SM contribution especially when the 
mixings are $(\delta^d_{LL})_{23} \sim (\delta^d_{RR})_{23} \sim 0.1$. For instance, we obtain 
$h_s = 1.8$ for $|(\delta^d_{LL})_{23}| = |(\delta^d_{RR})_{23}| \simeq 0.08$ in the case of 
$m_{\tilde q} = m_{\tilde g} = 500$GeV. 

As was mentioned in the previous section, the like-sign dimuon charge asymmetry favors a large 
$B_s - \bar{B}_s$ mixing phase. When the $a_2$ term dominates the SUSY contributions in 
(\ref{eq:b-bbar}), the following angle gives the CP phase, 
\begin{equation}
 \Theta^{+} = {\arg}(\delta^d_{LL})_{23} + \arg(\delta^d_{RR})_{23},
 \label{eq:theta+}
\end{equation}
where $\sigma_s$ is related to $\Theta^{+}$ as $\Theta^+ = 2\sigma_s -180^{\circ}$, 
e.g.~$\Theta^+ \simeq 20^{\circ}$ for $\sigma_s = 100^\circ$. 

Before proceeding to the experimental constraints, let us comment on the SUSY contribution to 
$\Gamma_{12}^s$. Although the dimuon anomaly implies a larger $\Gamma_{12}^s$ than the 
SM value~\cite{Ligeti:2010ia,Bauer:2010dga}, it is unlikely to expect this fulfilled in the SUSY 
models~(see \cite{Bauer:2010dga}). 

The large squark mixings are restricted by the other observables. In order to obtain the $\chi^2$ 
minimum (\ref{eq:dimuon}), the processes are considered which are sensitive to the new physics 
contributions to the $B_s - \bar{B}_s$ mixing. Moreover, the squark mixings $(\delta^d)_{23}$ 
can significantly contribute to the following observables:
\begin{enumerate}
 \item the branching ratio of the inclusive $b \to s \gamma$ decay: ${\rm Br}(b\to s \gamma)$,
 \item the time-dependent CP asymmetry of the $B_d$ decay into $\phi K$ and $\eta' K$: 
 $S_{\phi K_S}, S_{\eta' K_S}$,
 \item the atomic electric dipole moments through the strange quark CEDM: $d^c_s$.
\end{enumerate}
Note that their experimental result agrees with the SM predictions. Thus, they can give additional
constraint on $(\delta^d_{LL})_{23}$ and $(\delta^d_{RR})_{23}$.

Let us start from ${\rm Br}(b\to s \gamma)$. The experimental result~\cite{Barberio:2008fa} 
is consistent with the SM prediction~\cite{Misiak:2006zs}. In fact, the difference is limited in
\begin{equation}
  -0.3 \times 10^{-4} < \Delta {\rm Br}(b\to s\gamma) < 1.4 \times 10^{-4}.
\end{equation}
at the 2$\sigma$ level. In terms of the effective Hamiltonian, the branching ratio is approximately 
given by $|C_{7\gamma}|^2 + |\tilde C_{7\gamma}|^2$, where $C_{7\gamma}$ is the Wilson coefficient 
of the photonic magnetic operator, $O_{7\gamma} = \frac{e}{16\pi^2}m_b(\bar{s}_i\sigma^{\mu\nu}
P_Rb_i)F_{\mu\nu}$. It is known that the dominant SUSY contribution to $C_{7\gamma}$ behaves as
\begin{equation}
  (C_{7\gamma})_{\rm SUSY} \propto 
  (\delta_{LL}^d)_{23} \tan\beta\,
  m_{\rm soft}^{-2},
  \label{eq:bsg-SUSY}
\end{equation}
where $m_{\rm soft}$ is a typical soft mass scale, and $L \leftrightarrow R$ for $\tilde C_{7\gamma}$ 
(see \cite{Endo:2004dc} for details). 

In Fig.~\ref{fig:small_msoft}, we display the region which is allowed by ${\rm Br}(b\to s \gamma)$.
We want to mention that the analysis bases on $m_{\rm soft}=500$GeV and $\tan\beta = 10$. It is 
found that compared with the SUSY contributions to the $B_s-\bar B_s$ mixing (\ref{eq:b-bbar}), 
those to ${\rm Br}(b\to s \gamma)$ (\ref{eq:bsg-SUSY}) is enhanced when $\tan\beta$ is large 
and/or soft masses are small. Thus, the $b \to s \gamma$ constraint can exclude $h_s \sim 1$ 
for large $\tan\beta$ and small soft masses.
\footnote{ This constraint may be relaxed by a hierarchical squark mass spectrum between the first 
two and the third generations~\cite{Endo:2010fk}. }

It is noticed that since the SM contributes mainly to the Wilson coefficients $C_i$ (not $\tilde C_i$), 
the decay rate is sensitive to extra contributions to the real component of $C_i$. There are briefly
two contributions. First, the real component of $(\delta^d_{LL})_{23}$ can be finite with satisfying 
(\ref{eq:dimuon}), which induces ${\rm Re}C_{7\gamma}$. Secondly, the chargino, neutralino and 
charged Higgs diagrams can also contribute. In particular, those with the CKM matrix as a source 
of the flavor mixing interfere with the SM contribution. In Fig.~\ref{fig:small_msoft}, we take account 
of the chargino and charged Higgs diagrams with $m_{\rm soft}=500$GeV. They can change 
depending on the detailed mass spectrum of the superparticles, and we checked that there is a 
wide parameter region where (\ref{eq:dimuon}) is fulfilled with the constraint from ${\rm Br}(b\to s 
\gamma)$ satisfied. 

Secondly, the time-dependent CP asymmetries of the $B_d$ decays provide another constraint 
especially on the imaginary component of the squark mixings. In particular, those of $B_d \to \phi K$ 
and $B_d \to \eta' K$ have been measured well in the B-factories as~\cite{Barberio:2008fa}
\begin{equation}
 0.20 < S_{\phi K} < 0.88,~~~~~
 0.45 < S_{\eta' K} < 0.73.
\end{equation}
at the 2$\sigma$ level. If there is no NP contribution, it is considered that they become equal to the 
result of $B_d \to J/\psi K$, $S_{\psi K} = 0.67 \pm 0.2$~\cite{Barberio:2008fa}. In the SUSY models, 
they receive new contributions mainly through $C_{8G}$ and $\tilde C_{8G}$, which are the Wilson 
coefficients of the chromo magnetic operators. Similarly to (\ref{eq:bsg-SUSY}), the dominant SUSY 
contribution satisfies~\cite{Endo:2004dc}
\begin{equation}
  (C_{8G})_{\rm SUSY} \propto 
  (\delta_{LL}^d)_{23} \tan\beta\,
  m_{\rm soft}^{-2},
  \label{eq:cg8-SUSY}
\end{equation}
and $L \leftrightarrow R$ for $\tilde C_{8G}$. In the analysis, we use the generalized factorization 
method \cite{Ali:1997nh,Ali:1998eb} to evaluate the chromo-magnetic contribution to the decay 
amplitude, where the momentum which is transferred by the gluon of $O_{8G}$ is set $q^2 \simeq 
M_B^2/2$.

It is stressed that $S_{\phi K}$ and $S_{\eta' K}$ have different dependence on 
$(\delta^d_{LL})_{23}$ and $(\delta^d_{RR})_{23}$. Since the final state $\phi K$ has an odd parity, 
the decay amplitude depends on $C_i + \tilde C_i$, while noting that $\eta' K$ is parity even, the 
amplitude is proportional to $C_i - \tilde C_i$ (see \cite{Endo:2004dc}). As a result, we obtain the 
allowed region in Fig.~\ref{fig:small_msoft}. We notice that $S_{\eta' K}$ indicates $(\delta^d_{LL})_{23} 
\sim (\delta^d_{RR})_{23}$, while $S_{\phi K}$ does not give a stringent bound because of the 
large experimental uncertainty. 

Compared with the SUSY contributions to the $B_s-\bar B_s$ mixing (\ref{eq:b-bbar}), those to 
$(C_{8G})$ and $(\tilde C_{8G})$ increase when $\tan\beta$ is large and/or superparticles are 
light (see (\ref{eq:cg8-SUSY})). Thus, large $\tan\beta$ and/or small $m_{\rm soft}$ can close the 
window to realize $h_s \sim 1$. It is also commented that they are less sensitive to the real component 
of $C_{8G}$ and $\tilde C_{8G}$, because the real component of the decay amplitude is dominated 
by the SM contribution.

As a result of the above constraints from $b\to s \gamma$, $S_{\phi K}$ and $S_{\eta' K}$ as well as 
those sensitive to the anomalous $B_s - \bar{B}_s$ mixing, it seems that the SUSY contribution to the 
$B_s - \bar{B}_s$ mixing can be as large as the SM contribution with a large CP phase if the squark 
mixings are $(\delta^d_{LL})_{23} \sim (\delta^d_{RR})_{23} \sim 0.1$ for $m_{\rm soft} = O(100)$GeV 
and moderate $\tan\beta$. Moreover, focusing on the dependence on $(\delta^d_{LL, RR})_{23}$ and 
$m_{\rm soft}$ of $h_s$ and the constraints, a larger superparticle mass alleviates the constraints.

\begin{figure}[t!]
\begin{center}
\includegraphics{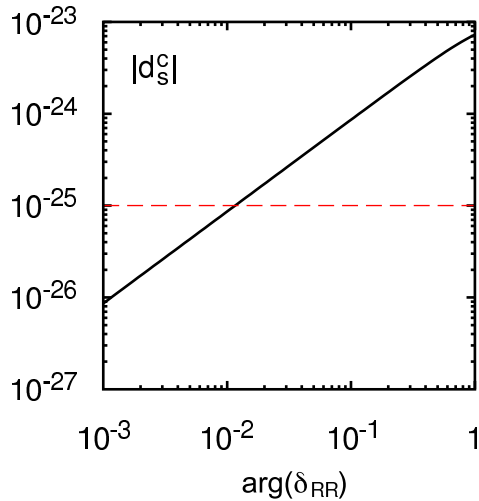}
\end{center}
\caption{
$|d_s^c|$ as a function of $\arg(\delta^d_{RR})_{23}$ with ${\rm arg}(\delta^d_{LL})_{23}=0$. 
The other parameters are $m_{\tilde{q}}=m_{\tilde{g}}=\mu=500\GEV$, $|(\delta^d_{LL})_{23}|=
|(\delta^d_{RR})_{23}|=0.08$ and $\tan\beta=10$. The red line shows $e|d_s^c| = 1.0 \times 
10^{-25} e{\rm cm}$ (see (\ref{eq:edm-bound})).}
\label{fig:edm_gluino}
\end{figure}

However, it is too early to conclude that the $B_s - \bar{B}_s$ mixing phase can be large with satisfying 
all the experimental constraints. The atomic and neutron EDMs are sensitive to the CP-violating phase 
of the $b-s$ squark mixings through the CEDM of the strange quark~\cite{Hisano:2003iw}. Based on the 
hadronic calculation and the experimental results~\cite{Baker:2006ts}, the strange quark CEDM is 
bounded as (see \cite{Hisano:2004tf})
\begin{equation}
  e|d_s^c| \lesssim 1\times 10^{-25} e{\rm cm}.
  \label{eq:edm-bound}
\end{equation}
On the other hand, if both $(\delta^d_{LL})_{23}$ and $(\delta^d_{RR})_{23}$ are finite, $d_s^c$ is 
dominated by the gluino--squark diagrams, and we obtain~\cite{Hisano:2003iw}. 
\begin{equation}
 e|d_s^c| 
  \propto |(\delta^d_{LL})_{23} (\delta^d_{RR})_{23}|
  \sin\Theta^{-} 
  \tan\beta\,
  m_{\rm soft}^{-2}
  \label{eq:edm}
\end{equation}
where $\Theta^-$ denotes the CP-violating phase, which is defined as
\begin{equation}
  \Theta^- = {\arg}(\delta^d_{LL})_{23}-\arg({\delta}^d_{RR})_{23}.
  \label{eq:edm-phase}
\end{equation}
Although $\Theta^-$ should also depend on the phase of the gluino mass and the $\mu$ 
parameter, they are tightly limited to be zero by the EDMs. Thus, we hereafter focus on 
the phase of $(\delta^d_{LL})_{23}$ and $(\delta^d_{RR})_{23}$. 

The EDMs provide a very severe constraint on the phase $\Theta^-$. In Fig.~\ref{fig:edm_gluino}, 
$|d_s^c|$ is evaluated as a function of $\arg({\delta}^d_{RR})_{23}$ with ${\arg}(\delta^d_{LL})_{23} 
= 0$, i.e.~$\Theta^- = -\arg({\delta}^d_{RR})_{23}$. Here, we choose the size of the squark 
mixings as $|(\delta^d_{LL})_{23}|=|(\delta^d_{RR})_{23}|=0.08$, which realize $h_s = 1.8$. 
Therefore, in order to fulfill (\ref{eq:dimuon}) ${\arg}(\delta^d_{LL})_{23}$ must be correlated with 
$\arg({\delta}^d_{RR})_{23}$ at the percent level. Even if we allow the CEDM bound up to 
$1 \times 10^{-24} e{\rm cm}$ due to a potentially large hadronic uncertainty, the 
phase is still required to tuned within the 10\% level. Thus, we find that it is unlikely to expect 
a large CP asymmetry in the $B_s$ meson mixing unless the phases satisfy $\Theta^+ \gg 
\Theta^-$ (see also \cite{Datta:2009fk}).

It is emphasized that the EDM constraint cannot be avoided even if we increase superparticle 
masses. This feature is contrasted to the constraints from $b\to s \gamma$, $S_{\phi K}$ and 
$S_{\eta' K}$. This is because the gluino--squark contribution to $d_s^c$ (\ref{eq:edm}) is 
proportional to $(\delta^d)_{23}^2 m_{\rm soft}^{-2}$, which is the same structure as $h_s$ 
(\ref{eq:b-bbar}). This means that the EDMs provide more robust constraint on the $B_s-\bar B_s$ 
mixing phase than $b\to s \gamma$, $S_{\phi K}$ and $S_{\eta' K}$, and it is considered that 
the $B_s-\bar B_s$ mixing phase is difficult to be large. 

The strict EDM bound may be reconciled with the large $B_s - \bar B_s$ mixing phase by supposing 
cancellations among the SUSY contributions to the EDMs. In fact, there are additional contributions 
such as those from the neutralino and the chargino, which can interfere with the gluino contribution. 
Moreover, the SUSY contribution to the $B_s - \bar B_s$ mixing phase, $\Theta^+$, in (\ref{eq:theta+}) 
is not equal to the CP-violating phase appearing in the CEDM, $\Theta^-$, in (\ref{eq:edm-phase}).
Thus, the CEDM bound can be avoided if $\Theta^-$ vanishes, i.e.~${\arg}(\delta^d_{LL})_{23} 
\simeq \arg({\delta}^d_{RR})_{23}$, while the $B_s - \bar B_s$ mixing phase is left sizable as long as 
${\arg}(\delta^d_{LL})_{23}$ and $\arg({\delta}^d_{RR})_{23}$ are large~\cite{Endo:2010fk}. 

This (accidental) cancellation is, however, easily polluted by radiative corrections. One of the 
most ubiquitous corrections is renormalization group contributions to the soft parameters. 
Even if we suppose $(\delta^d_{LL})_{23} = (\delta^d_{RR})_{23}$ at a high energy scale, 
$(\delta^d_{LL})_{23}$ receives a correction during the evolution down to the weak scale as
\begin{equation}
  \Delta \left(\delta^d_{LL}\right)_{23} \simeq
  -\frac{1}{2\pi^2} V_{ts}^* Y_t^2 \ln\frac{M_{X}}{M_{\rm SUSY}},
\end{equation}
where $M_X$ and $M_{\rm SUSY}$ are the input and SUSY scales, respectively. For instance, 
if $M_X$ is the GUT scale, $\Delta \left(\delta^d_{LL}\right)_{23}$ becomes $O(10^{-2})$. It is 
emphasized that the phase of the correction is aligned to the CKM phase, that is real in this article. 
Thus, this generally shifts the total phase of $(\delta^d_{LL})_{23}$ by $O(0.1-1)$ when the mixing 
is $|(\delta^d_{LL})_{23}| \sim (0.01-0.1)$. This means that it is quite unlikely to expect the cancellation 
among the SUSY contributions to the CEDM unless it is protected by symmetries. 
\footnote{When the CP violation originates from the Yukawa couplings apart from the CKM matrix, 
the hermicity of the Yukawa matrix ensures the suppression of the EDMs in relation to the strong 
CP problem~\cite{Endo:2010fk}. }
Consequently, it is very difficult to reconcile the large $B_s - \bar B_s$ mixing phase with the 
EDM bound when the gluino diagrams dominate the SUSY contributions. 

\section{Decoupling Case}

When the gluino is relatively light, we showed in the previous section that the large CP asymmetry 
in the $B_s - \bar{B}_s$ mixing conflicts with the EDM bound generically. In order to avoid the 
constraint, let us consider heavy superparticle scenarios. When the colored superparticles are 
sufficiently heavy, the gluino--squark contributions are suppressed. 

There are possibly sizable contributions from the heavy Higgs bosons as long as their masses 
are $O(100-1000) \GEV$. It is stressed that the Higgs contributions are enhanced when $\tan\beta$ 
is large. The Lagrangian of the Higgs bosons and the down-type quarks is written as~\cite{Hall:1993gn,Carena:1994bv}
\begin{equation}
\mathcal{L} = Y_i Q_i H_d D^c_i  + Y'_{ij} Q_i H_u^* D^c_j + {\rm h.c.},
\end{equation}
where $Y_i$ is the Yukawa coupling of the down-type quark, $i$, and $Y'_{ij}$ is a non-holomorphic 
Yukawa coupling, which is generally complex and flavor non-diagonal. Diagonalizing the down-type 
quark mass matrix, we obtain the flavor-changing (and CP-violating) neutral Higgs couplings as~\cite{Hall:1993gn,Carena:1994bv,Isidori:2001fv}
\begin{equation}
\mathcal{L} = Y'_{ij} Q_i H_u^* D^c_j - Y'_{ij} \tan\beta\, Q_i D^c_j H_d + {\rm h.c.}.
\label{eq:non-holomorphic}
\end{equation}
It is noticed that the Higgs bosons couple to the down-type quarks with flavor violations due to the 
non-holomorphic interaction. 

Within the framework of SUSY, the non-holomorphic interactions are not involved in the Lagrangian 
at the tree level. Rather, they are generated by radiative corrections to the Higgs coupling with the 
matter fermions, whose diagrams include the superparticles~\cite{Hall:1993gn,Carena:1994bv}. 
The important features of them are that the non-holomorphic Yukawa couplings are proportional to 
$\tan\beta$ and are not suppressed when the superparticles decouple. As a result, the second term 
in (\ref{eq:non-holomorphic}) is proportional to $\tan^2\beta$ even in the decoupling limit of the 
superparticles. It is also noted that the heavy Higgs bosons mainly consist of $H_d$ when their 
masses are relatively large compared to the electroweak scale. Thus, the diagrams intermediating 
the heavy Higgs bosons are sizably enhanced when $\tan\beta$ is large, even though the couplings 
emerges at the one-loop order.

\begin{figure}[thbp]
\begin{center}
\includegraphics[width=15.0cm]{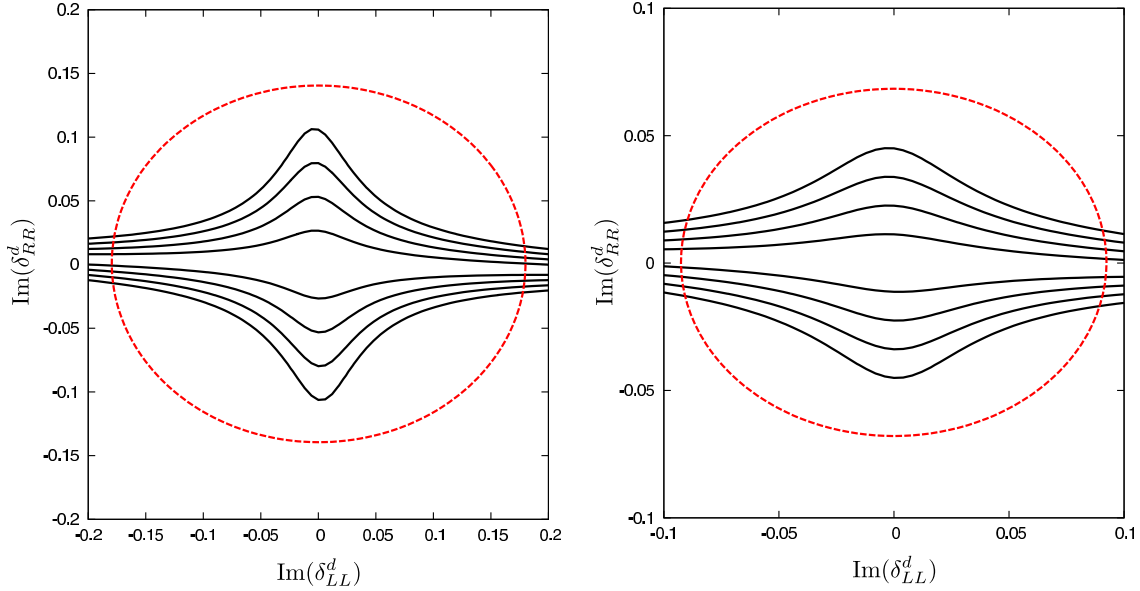}
\end{center}
\caption{Contours of $h_s=2.0,1.5,1.0$ and $0.5$ from the outside to inside. The constraint 
from ${\rm Br}(B\to \mu^+ \mu^-)$ is shown by the dashed (red) line. We take $m_H = 500$GeV 
and ${\rm Re}(\delta_{LL}^d)={\rm Re}(\delta_{RR}^d)=0$. In the left (right) panel, we set $\tan\beta
=30 (40)$.}
\label{fig:cont_large_msusy}
\end{figure}
\begin{figure}[thbp]
\begin{center}
\includegraphics[width=7.5cm]{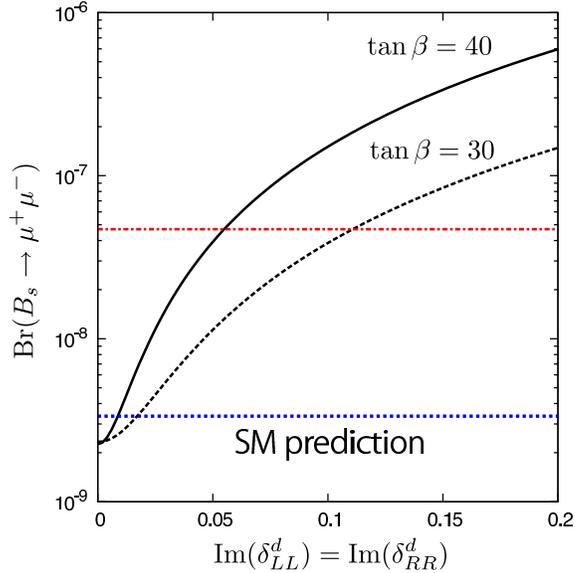}
\end{center}
\caption{
$\BR(B_s \to \mu^+ \mu^-)$ for $\tan\beta = 30$ and 40, where the squark mixings are varied with 
${\rm Im}(d^d_{LL})_{23}={\rm Im}(d^d_{RR})_{23}$ fixed. The parameters are ${\rm Re}
(\delta_{LL}^d)_{23}={\rm Re}(\delta_{RR}^d)_{23}=0$ and $m_H=500 \GEV$. The dash-dotted (red) 
line shows the current experimental bound, $\BR(B_s\to\mu^+\mu^-)=4.7\times 10^{-8}$~\cite{Nakamura:2010zzi}, and the dotted-line (blue) is the SM prediction.}
\label{fig:bsmm}
\end{figure}

The neutral heavy Higgs diagram can enhance the $B_s - \bar{B}_s$ mixing. The double Higgs 
penguin diagrams dominantly contribute to $M_{12}^s$, which include a couple of the heavy 
Higgs couplings. Then, $M_{12}^s$ is roughly enhanced by $\tan^4\beta$. 
The contribution becomes~\cite{Foster:2005wb}
\footnote{In the numerical analysis, we also include the charged Higgs 
contributions~\cite{Foster:2005wb}, though they are subdominant compared 
to the neutral Higgs contributions.}
\begin{equation}
 h_s e^{2i\sigma_s} \simeq \Bigr[a'_1 V_{ts}^* (\delta^d_{RR})_{23} - a'_2 (\delta^d_{LL})_{23}
 (\delta^d_{RR})_{23} \Bigl], 
 \label{eq:higgs_hs}
\end{equation}
where we explicitly show the flavor mixings which come from the non-holomorphic couplings. 
In particular, the first term in the right-hand side consists of the chargino and gluino contributions 
to the non-holomorphic couplings, and the second term originates from the gluinos. The coefficients 
$a'_i$ include the loop functions, and they are estimated to be a real and positive number. 
To be explicit, they depend on parameters as
\begin{equation}
 a'_i \propto \frac{\tan^4\beta}{(1+\epsilon\tan\beta)^4} \frac{1}{m_H^2},
\end{equation}
where the heavy Higgs mass, $m_H$, arises from the contributions from the CP-even and CP-odd 
heavy Higgs bosons. Quantitatively, it is estimated as $a'_1 \sim a'_2$. In contrast to the gluino 
dominant case, we notice that $h_s$ can be sizable solely by $(\delta^d_{RR})_{23}$, while the 
contribution which depends only on $(\delta^d_{LL})_{23}$ is suppressed by a chiral factor 
$m_s/m_b$. Therefore, a large $h_s$ favores large $(\delta^d_{RR})_{23}$. 

The constant $\epsilon$ comes from the non-holomorphic corrections to the Yukawa couplings of 
the down-type quarks. It satisfies $Y'_{ii} \sim Y_i \epsilon$, and $\epsilon$ is not enhanced by 
$\tan\beta$ (see \cite{Foster:2005wb} for details). When superparticles are almost degenerate, it is 
estimated as $|\epsilon| \sim 0.01$, and its sign is determined by the sign of $\mu$ as ${\rm sgn}
(\epsilon)={\rm sgn}(\mu)$. As we will explain later, since negative $\mu$ leads to larger $d_s^c$ 
and $\BR(b \to s\gamma)$, positive sign is preferred. In this article we take $\mu>0$, if not otherwise 
mentioned. 

We shows contours of $h_s$ in Fig.~\ref{fig:cont_large_msusy}, where we take 
${\rm Re}(\delta^d_{LL})_{23}={\rm Re}(\delta^d_{RR})_{23}=0$ and $m_{H}=500\GEV$. We also 
choose two cases of $\tan\beta$ as $\tan\beta=30$ and 40 in the left and right panels, respectively. 
It is found that the heavy Higgs contributions to the $B_s - \bar{B}_s$ mixing become comparable to 
the SM contribution for $(\delta^d_{LL,RR})_{23} \sim 0.1$. For example, we obtain $h_s = 1.8$ when 
the mixings are ${\rm Im}(\delta^d_{RR})_{23} \simeq 0.07 (0.03)$ for $\tan\beta=30 (40)$. 

The flavor-changing neutral Higgs couplings also enhance the branching ratio of the $B_s \to \mu^+ 
\mu^-$ decay, significantly. The dominant contribution to the decay amplitude includes the 
non-holomorphic interactions and the Yukawa coupling of the muon. Thus, $\BR(B_s \to \mu^+ \mu^-)$ 
increases at $\tan^6\beta$. In terms of the effective Hamiltonian, the branching ratio is represented 
by the Wilson coefficients as~\cite{Foster:2005wb}
\begin{equation}
 \BR(B_s \to \mu^+ \mu^-) \propto 
 \left[0.9 |C_S -\tilde{C}_S|^2 + |C_S +\tilde{C}_S-A_{SM}|^2\right],
\end{equation}
where $A_{SM}$ denotes the SM contribution. The Wilson coefficients $C_S$ and $\tilde{C}_S$ are 
induced by the heavy Higgs bosons and given as
\begin{eqnarray}
 C_S &\propto& \frac{\tan^3\beta}{(1+\epsilon\tan\beta)^2}
 \frac{1}{m_H^2} \left[ {\rm sgn}(A_t) V_{tb} V_{ts}^* -a \,(\delta^d_{LL})_{23} + 
 b \,(\delta^d_{RR})_{23} \right], \nn
 \tilde{C}_S &\propto& 
 \frac{\tan^3\beta}{(1+\epsilon\tan\beta)^2} \frac{1}{m_H^2} 
 \left[b  \, (\delta^d_{RR})_{23}\right].
\end{eqnarray}
In the bracket, we explicitly show the flavor mixings and a relative size of the contributions by the 
coefficients, $a$ and $b$, which are estimated as $a \sim0.1$ and $b \sim 1$, respectively. 
Numerically, $C_S$ and $\tilde{C}_S$ exceed $A_{SM}$ when the squark mixings are 
$(\delta^d_{LL,RR})_{23} \gsim 0.1$. Then, $\BR(B_s \to \mu^+ \mu^-)$ is roughly proportional 
to $|(\delta^d_{LL})_{23}|^2 + |(\delta^d_{RR})_{23}|^2$.

The $B_s \to \mu^+ \mu^-$ decay has not been measured yet, and the branching ratio is constrained 
by the Tevatron experiments as
\begin{equation}
 \BR(B_s \to \mu^+ \mu^-) < 4.7 \times 10^{-8}. \label{eq:br_bsmu2}
\end{equation}
at the 90\% C.L.~\cite{Nakamura:2010zzi}. In Fig.~\ref{fig:cont_large_msusy}, we show the constraint 
of $\BR(B_s \to \mu^+ \mu^-)$. Namely, the region inside the ellipse is consistent with the experiment. 
We notice that the contribution proportional to $(\delta^d_{LL})_{23}$ becomes smaller than those to 
$(\delta^d_{RR})_{23}$, because the sign of $a$ is opposite to that of $b$ in $C_S$. Therefore, the 
allowed region for $(\delta^d_{LL})_{23}$ is rather large in Fig.~\ref{fig:cont_large_msusy}. Anyway, 
we obtain a large $B_s-\bar{B}_s$ mixing without conflicting with the constraint of $\BR(B_s \to \mu^+ 
\mu^-)$ in a wide parameter region.

In the LHCb experiment, it is expected that the sensitivity of $\BR(B_s \to \mu^+ \mu^-)$ reaches 
the SM prediction\cite{Bettler:2009rf}, which is $\BR(B_s \to \mu^+ \mu^-)=(3.35 \pm 0.32) \times 
10^{-9}$~\cite{Blanke:2006ig}. In Fig.~\ref{fig:bsmm}, $\BR(B_s \to \mu^+ \mu^-)$ is shown as a
function of the size of the mixing with ${\rm Im}(\delta^d_{LL})_{23}={\rm Im}(\delta^d_{RR})_{23}$ 
fixed. When $(\delta^d_{LL,RR})_{23}$ is large enough to satisfy (\ref{eq:dimuon}), the branching 
ratio is found to be close to the current upper bound for $\tan\beta=40$. It is noted that $\BR(B_s 
\to \mu^+ \mu^-)$ depends on the heavy Higgs boson mass. We set the mass to be 500GeV in 
Fig.~\ref{fig:bsmm}. Compared with the dependence of $h_s$ on $m_H$ and $(\delta^d)_{23}$, 
$\BR(B_s \to \mu^+ \mu^-)$ relatively decreases as $m_H$ becomes larger with $h_s$ fixed. 
Nonetheless, $\BR(B_s \to \mu^+ \mu^-)$ tends to be sizable to obtain a large $B_s-\bar B_s$ 
mixing phase, and such a large branching ratio is expected to be measured in the LHCb. 
\footnote{Note that since the contributions from the real part of $(\delta^d)_{23}$ interferes with SM
and minimal flavor-violating contributions, $\BR(B_s \to \mu^+ \mu^-)$ can be affected by the real 
component.}

Let us turn to the experimental constraints from the EDMs, $\BR(b \to s\gamma)$, $S_{\eta' K_S}$ 
and $S_{\phi K_S}$. In the gluino dominant case, the EDMs provide especially tight constraints on 
the CP violations. In contrast, when the superparticles are heavy, we will see that these constraints 
are greatly ameliorated. 

In the decoupling limit, these processes are induced by the charged Higgs diagrams. Similarly to 
the heavy neutral Higgs bosons, the flavor-changing (CP-violating) charged Higgs contributions 
are generated by the non-holomorphic interactions. Since the mass of the charged Higgs boson, 
$m_{H^\pm}$, is likely to degenerate with that of the heavy Higgs bosons up to SU(2) breaking 
corrections, we hereafter assume $m_{H^\pm} = m_H$ for simplicity. 

\begin{figure}[tbp]
\begin{center}
\includegraphics[width=7.5cm]{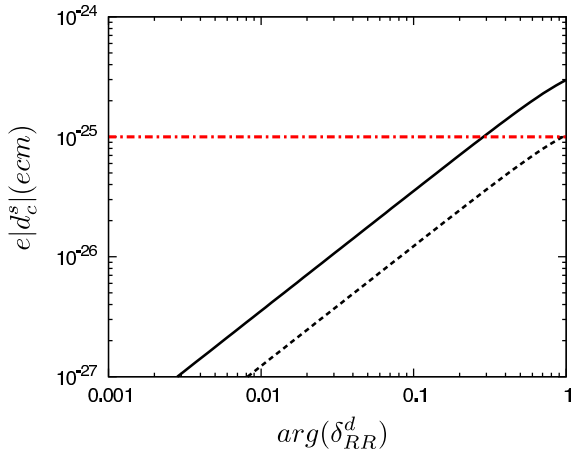}
\end{center}
\caption{$d_s^c$ as a function of $\arg(\delta^d_{RR})_{23}$. We take $\arg(\delta^d_{LL})_{23}=0$ 
and $m_{H^\pm}=500 \GEV$. The solid (dashed) line corresponds to $|(\delta^d_{LL})_{23}|=
|(\delta^d_{RR})_{23}|=0.07(0.03)$ for $\tan\beta=40(30)$. The dash-dotted (red) line shows 
$e|d_s^c|=1.0\times 10^{-25}e$cm.}
\label{fig:edm_large_msusy}
\end{figure}

The charged Higgs contribution to the CEDM of the strange quark is given at the BLO 
as~\cite{Hisano:2008hn}. 
\begin{equation}
 e d_s^c \propto \frac{\tan\beta}{(1+\epsilon\tan\beta)^2} \, 
 {\rm Im}\left[ V_{ts} (\delta^{d}_{RR})_{23} \right] \frac{1}{m_{H^\pm}^2},
 \label{eq:edm_largemsusy}
\end{equation}
where the CKM matrix gives arise at the charged Higgs vertex, and the squark mixing originates 
in the the non-holomorphic Yukawa coupling which is dominated by the gluino diagrams. In addition, 
there is a contribution which depends on $[ (\delta^{d}_{LL})_{23} (\delta^{d}_{RR})_{23} ]$. This 
term is obtained by substituting the vertex of the CKM matrix into another non-holomorphic coupling. 
Since the non-holomorphic interaction is smaller than the top Yukawa coupling, its contribution to 
$d_s^c$ is smaller by one order of magnitude than (\ref{eq:edm_largemsusy}).
Thus, the EDM is sensitive to the phase of $(\delta^{d}_{RR})_{23}$. Although the neutral Higgs 
bosons can also contribute to $d_s^c$, the effects are negligible because the correction from the 
CP-even heavy Higgs boson cancels with that from the CP-odd heavy Higgs boson when the mass 
splitting between $m_{A}$ and $m_{H}$ is small~\cite{Hisano:2008hn}. 

Fig.~\ref{fig:edm_large_msusy} shows the dependence of $d_s^c$ on the phase of 
$(\delta^d_{RR})_{23}$ in the cases of $\tan\beta=30$ and 40. In the analysis, the mixings 
are chosen as $(\delta^d_{RR})_{23}=0.07$ and $0.03$, respectively, which are required to 
realize $h_s = 1.8$. It is emphasized that $d^s_c$ is proportional to $(\delta^q)_{23} \tan\beta$. 
This is contrasted to the SUSY contributions to the $B - \bar B$ mixing and $\BR(B_s \to \mu^+ \mu^-)$, 
which briefly depend on $(\delta^q)_{23}^2 \tan^4\beta$ and $(\delta^q)_{23}^2 \tan^6\beta$, respectively. 
Moreover, $d_s^c$ is scaled by $m_{H^\pm}^{-2}$, which is the same as $h_s$. Thus, when $h_s$ is 
given, the EDM bound is satisfied especially when $\tan\beta$ and $m_{H^\pm}$ are larger. Actually, 
for $\tan\beta=40$ and $(\delta^d_{RR})_{23}=0.03$, $e |d_s^c|$ is smaller than $1.0 \times 10^{-25}
e {\rm cm}$ even with $\arg(\delta^d_{RR})_{23} = O(1)$. Then, $\BR(B_s \to \mu^+\mu^-)$ is likely 
to be large in the region.

The charged Higgs contribution to $d_s^c$ is enhanced when $\epsilon$ is negative due to the factor 
$(1+\epsilon\tan\beta)^{-2}$ in (\ref{eq:edm_largemsusy}). As is the case of $h_s$, it comes from 
the non-holomorphic corrections to the Yukawa couplings of the down-type quarks. Note that the sign 
of $\epsilon$ is equal to that of $\mu$. Hence, when $\mu$ is negative, $d^s_c$ becomes larger by 
$O(10)$ than the result of positive $\mu$. As a result, $d^s_c$ can exceed the experimental bound 
for $\arg(\delta^d_{RR})_{23} = O(1)$. Thus, $\mu > 0$ is preferred by the EDMs. 
 
The charged Higgs boson also contributes to $\BR(b \to s\gamma)$, $S_{\eta' K_S}$ and 
$S_{\phi K_S}$ through $C_{7\gamma}$ and $C_{8G}$. At the BLO, $C_{7\gamma}$ and 
$C_{8G}$ become~\cite{Foster:2005wb}
\begin{eqnarray}
 C_{7\gamma, 8G} &\propto& 
 \frac{1-\epsilon\tan\beta}{1+\epsilon\tan\beta} \frac{1}{m_{H^\pm}^2} + 
 \mathcal{O}(10^{-1}) \times 
 \frac{\tan\beta}{1+\epsilon\tan\beta}(\delta^d_{LL})_{23} \frac{1}{m_{H^\pm}^2}, \nn
 \tilde{C}_{7\gamma,8G} &\propto& \mathcal{O}(10^{-1}) \times 
 \frac{\tan\beta}{1+\epsilon\tan\beta} (\delta^d_{RR})_{23}\frac{1}{m_{H^\pm}^2}, 
 \label{eq:c78lt}
\end{eqnarray}
where the first term in the right-hand side of $C_{7\gamma,8G}$ comes from the charged Higgs 
diagram with the CKM matrix. On the other hand, the flavor mixing in the second term and that of 
$\tilde C_{7\gamma,8G}$ originate from the non-holomorphic interaction which is dominated by 
the gluino diagrams.

Although we omit the numerical coefficients, it is checked that (\ref{eq:c78lt}) is quantitatively 
smaller than the SM contribution for $\mu>0$. Actually, the first term of $C_{7\gamma,8G}$ 
becomes $O(10^{-2})$ for $m_{H^\pm}=500 \GEV$ and $\tan\beta=30-40$, which is just a small 
correction to the SM value, $(C_{7\gamma,8G})_{SM} \sim 0.1$. On the other hand, the other 
terms are linearly proportional to $\tan\beta$, in contrast to $h_s$ and $\BR(B_s \to \mu^+\mu^-)$. 
Numerically, we obtain $C_{7\gamma,8G} \sim 10^{-2}$ for $\tan\beta = O(10)$ and $(\delta^d)_{23} 
\sim 0.1$ from these terms. As a result, $\BR(b \to s\gamma)$, $S_{\phi K_S}$ and $S_{\eta' K_S}$ 
remain consistent with the experimental values.

When $\mu$ is negative, $C_{7\gamma, 8G}$ becomes larger that the case of $\mu>0$. Actually, 
they can reach ${\cal O}(10^{-1})$ for large $\tan\beta$ mainly from the first term in (\ref{eq:c78lt}). 
Since this is comparable to the SM values, the charged Higgs contributions to $\BR(b\to s\gamma)$ 
can exceed the experimental bound. On the other hand, noting that the first term in (\ref{eq:c78lt}) 
is real, i.e.~aligned to the SM contribution, this contribution to $S_{\eta' K_S}$ and $S_{\phi K_s}$ is
irrelevant. It is also checked that the other contributions in (\ref{eq:c78lt}) are less significant for 
$S_{\eta' K_S}$ and $S_{\phi K_s}$, because they are smaller than the first term.

\section{Conclusion}

Motivated by the like-sign dimuon charge asymmetry, we have investigated the CP asymmetry 
in the $B_s-\bar{B}_s$ mixing. Although a large mixing phase is consistent with the experimental 
constraints from ${\rm Br}(b \to s\gamma)$ and the mixing-induced CP asymmetries of the $B_d$ 
decays into $\phi K$ and $\eta' K$, we have shown that when the gluino diagrams dominate the 
SUSY contributions, the EDM bound almost excludes this possibility unless the cancellation works 
among the SUSY contributions to the EDM. 

In order to alleviate the tight EDM constraint, we have analyzed the heavy superparticle scenarios. 
In the decoupling limit of the superparticles except for the Higgs bosons, the FCNCs and CP violations 
are indued by intermediating the neutral/charged heavy Higgs bosons. We have found that the 
non-holomorphic Yukawa couplings can enhance the $B_s - \bar{B}_s$ mixing with satisfying the 
experimental constraints from the EDMs as well as ${\rm Br}(b \to s\gamma)$, $S_{\phi K}$ and 
$S_{\eta' K}$. In the region where the $B_s-\bar{B}_s$ mixing phase is large, 
$\BR(B_s \to \mu^+ \mu^-)$ tends to be sizable, and we expect to observe the decay at the LHCb.

\section*{Acknowledgements}
N.Y. is supported by Grand-in-Aid for Scientific Research, No.22-7585 from JSPS, Japan. 
This work was supported by World Premier International Research Center Initiative (WPI Initiative), 
MEXT, Japan.

\providecommand{\href}[2]{#2}\begingroup\raggedright\endgroup

\end{document}